\documentclass[AMA,STIX1COL]{WileyNJD-v2}

\articletype{Article Type}%

\received{date}
\revised{date}
\accepted{date}

\begin{document}

\title{Truffle tests for free -\\ Replaying Ethereum smart contracts for transparency}

\author[1]{Pieter Hartel}
\author[2]{Mark van Staalduinen}
\authormark{AUTHOR ONE \textsc{et al}}

\address[1]{\orgdiv{iTRust}, \orgname{Singapore University of Technology and Design}, \orgaddress{\country{Singapore}}, ORCID 0000-0002-0411-0421}
\address[2]{\orgdiv{TNO-SEA}, \orgname{The Netherlands Organisation for Applied Scientific Research}, \orgaddress{\country{Singapore}}, ORCID 0000-0001-9716-3815}

\corres{\email{pieter\_hartel@sutd.edu.sg} \\ \\
{\bf Funding} \\
This work was supported in part by the National Research Foundation (NRF), Prime Minister's Office, Singapore, under its National Cybersecurity R\&D Programme (Award No. NRF2016NCR-NCR002-028) and administered by the National Cybersecurity R\&D Directorate.}

\abstract[summary]{The Ethereum blockchain is essentially a globally replicated public database. Programs called smart contracts can access this database. Over 10 million smart contracts have been deployed on the Ethereum blockchain. Executing a method of a smart contract generates a transaction that is also stored on the blockchain. There are over 1 billion Ethereum transactions to date. Smart contracts that are transparent about their function are more successful than opaque contracts. We have therefore developed a tool (ContractVis) to explore the transparency of smart contracts. The tool generates a replay script for the historic transactions of a smart contract. The script executes the transactions with the same arguments as recorded on the blockchain, but in a minimal test environment. Running a replay script provides insights into the contract, and insights into the blockchain explorer that was used to retrieve the contract and its history. We provide five concrete recommendations for blockchain explorers like Etherscan to improve the transparency of smart contracts.}



\maketitle
\thispagestyle{plain}
\pagestyle{plain}

\section{Introduction}
\label{sec:introduction}
Software transparency is defined as the condition that all functions of software are disclosed to its users~\cite{Meunier2008}. Most software is complex, which makes it easy for developers to hide specific functions, either deliberately or inadvertently~\cite{Pfleeger2016}. This complexity reduces transparency. In this paper we study blockchain-related software, and in particular smart contracts on Ethereum, written in Solidity. Smart contracts on Ethereum are not normally complex~\cite{Hegedus2019}, which should improve their transparency. Ethereum smart contracts execute on a complex infrastructure, the backbone of which, the blockchain, is a globally replicated data structure that is available for public scrutiny. This should also contribute to transparency of the infrastructure~\cite{Fontana2018}.

One of the tools commonly used for the analysis of the public blockchain is the blockchain explorer. It interprets the blockchain and shows the complete history of every contract ever made. The most prominent blockchain explorer for Ethereum is Etherscan. In addition to the standard features of a blockchain explorer, such as listing blocks and transactions, Etherscan allows developers to upload the source code of a smart contract. Etherscan then verifies that the bytecode that is deployed on the blockchain can be reproduced by compiling the uploaded source. This allows the user of a verified smart contract to inspect the source code of the contract. But what if the user is not able to audit the source code? An alternative approach would be to deploy sophisticated program analysis tools to analyse the contract. Such approaches work with abstractions that must be proved sound. We discuss the merits of the most prominent approaches in Section~\ref{sec:related}.

Our approach to analyse contracts is a different one. We propose to ``Try the simplest thing first''. By this we mean executing the contract in a minimal testing environment, and then to compare the execution results to the historic execution results of the contract on the blockchain. We will work with the Solidity source of a smart contract. It would be possible to decompile something resembling source code from bytecode but this will never be as transparent as the original source because comments cannot be recovered by decompilation. We use standard tools to execute smart contracts, which means that soundness is not an issue. We have developed a tool that generates a test script for a verified smart contract from its historic transactions on the blockchain. If the contract fails, it must depend either on other contracts, or on the infrastructure. This is evidence of opacity. In theory it is possible to discover the dependencies of a smart contract by analysing the blockchain. However, current blockchain explorers do not disclose all dependencies. This made it necessary to develop ContractVis to discover (a) all dependencies, and (b) the extent to which dependencies hide functions of smart contracts, thereby reducing transparency. 

The tool can be used in three different ways. Firstly, when a test is executed in the minimal test environment, the historic transactions are replayed. If the contract does not have any dependencies on other contracts, or on the infrastructure, it will successfully execute. This is evidence of transparency. 

Secondly, when the contract, the test, and / or the configuration of the environment are changed prior to replaying, we can run experiments. For example we can explore what-if questions, such as what if the time stamp of the current block is a sequence (e.g. 1,2,3)? Would our lottery contract still be fair?

Thirdly, a generated test can be used as a regression test in further development of the contract, or as input to other testing tools, such as fuzzers and mutators.

Debuggers and fuzzers can also replay smart contracts. The Truffle development framework (See~\url{http://truffleframework.com}) represents the state-of-the-art in smart contract debugging. By specifying the unique hash of a transaction to the command {\normalsize \verb=truffle debug=} it is possible to replay an existing transaction on the blockchain, stepping into and over calls, single-stepping the execution of the bytecode etc. Other development environments offer similar facilities. However, debuggers usually do not generate test scripts that developers can experiment with, hence exploring what-if questions is infeasible with existing debuggers.

Fuzzers~\cite{Jiang2018} generate tests with well-chosen, random inputs to drive the contract into an error state. The developer can then use the information about the inputs and the error state to improve the contract. Our approach is different in that we use {\it historic} inputs to drive the contract into a correct state. If this fails and the contract reaches an error state, the developer has gained insight into the contract.

Smart contracts are unique in the sense that the complete set of historic inputs to each contract is available on the blockchain. For programs other than smart contracts, some logging information is usually available, but this is rarely sufficient to replay the program. For example checkpoint restart systems~\cite{Hargrove2006} take great care to augment logs with the minimum information necessary to replay an execution efficiently. Debugging systems for concurrent programs require significant sophistication to be able to reproduce concurrent executions faithfully~\cite{Choi1998}.

Smart contracts should not need extra information for accurate replay. After all, each full node that joins the Ethereum peer-to-peer network replays all historic transactions. The question is, to what extent the historic transactions of a smart contract can be replayed on a minimal testing environment. By a minimal testing environment we mean using (a) the source of smart contracts and historic transactions (i.e. Etherscan), (b) the testing framework to replay the contracts (i.e. the Truffle framework), and (c) bespoke scripting (i.e. ContractVis) to translate historic transactions into {\normalsize \verb=truffle test=} scripts.

Our contributions are as follows. Firstly, we measure to what extent smart contracts depend on other contracts and on the environment. To learn of all the functionalities of such contracts, users have to study not only the source, but also the dependencies. The more dependencies a contract has, the more effort will be required to achieve transparency. Secondly, we show that certain aspects of the behaviour of smart contracts can be tracked intuitively with automatically generated heat maps. Thirdly, we suggest improvements to current blockchain explorers that give users and developers automated insight into the dependencies. Finally, the paper contains details on the implementation of smart contracts on Ethereum that is available on the web but not in one place. We hope that by exposing this information we contribute to transparency.

The next section discusses the background and presents the research questions. Section~\ref{sec:method} describes the experimental approach and Section~\ref{sec:results} presents the results of the experiments. Recommendations for the developers of blockchain explorers are provided in Section~\ref{sec:recommendations}. This is followed by the discussion, related work, limitations, conclusions and future work.

\section{Background}
We summarise how smart contracts are executed, how they can be replayed and how the accuracy of the replay scripts can be measured. We then formulate a number of research questions.

\subsection{Smart contracts}
A {\it smart contract} is a program that executes and stores its data on a globally replicated database called a blockchain. The constructor and the methods of a smart contract encode the business logic of a distributed web application called a Distributed Application (DApp). The call to the constructor and every method call made by the DApp create one or more transactions, which are stored in a specific block on the blockchain. The blockchain is basically an append-only, replicated database. A global peer-to-peer network operated by so-called miners manages the Ethereum blockchain. The miners run a distributed agreement protocol to decide which blocks are append to the blockchain. The miners are financially rewarded for their efforts. Once stored in a block on the blockchain, transactions normally persist (except in the case of forks). Because of the global replication of the blockchain, transactions are censorship resistant.

When a contract is deployed, the Ethereum Virtual Machine (EVM) bytecode obtained by compiling the source of the contract is also stored on the blockchain. The blockchain does not only store contracts and transactions but also accounts with a monetary balance in Ether, which is the native currency of Ethereum. At the time of writing, the value of an Ether is about 330 US\$. There are two types of accounts: an externally owned account, defined by creation of a public private key pair and a contract account, defined by the deployment of a contract. Each transaction has a unique hash and each contract account or externally owned account is identified by a unique address.

When the constructor and the methods of a contract are executed, a sequence of {\it transactions} is generated, starting with the transaction that created the contract. Each transaction changes the state of the contract variables. A transaction may change the balance of accounts, and it usually emits events to pass information to the controlling DApp. A transaction does not have a return value. Every successful transaction consumes a certain amount of gas proportional to the amount of work performed by the contract. Gas costs Ether and the total gas consumption of a transaction is therefore an interesting output of a contract. A transaction may fail and it therefore has a status. For example, if the amount of gas provided is insufficient, the transaction will fail. This mechanism is used, amongst others, to prevent infinite loops. Transactions are cryptographically signed to ensure that only users in possession of the appropriate private key can make transactions.

Ethereum has two types of transactions: {\it internal and external}. The external transactions are triggered by externally owned accounts. Internal transactions are triggered by contract accounts.

Transactions can fail programmatically, via calls to either {\normalsize \verb=require()=} or {\normalsize \verb=assert()=}.
{\normalsize \verb=require()=} is used to check external consistency (e.g. to make sure that inputs satisfy specific requirements).
{\normalsize \verb=assert()=} is used to check internal consistency (e.g. to ensure that the balance of an account does not underflow).

A contract does not only contain methods that change the state but also {\it pure functions} that only read the state of the contract from the blockchain. Pure functions are called to return information about the state of the contract to the DApp. Calls to pure functions always succeed and do not consume gas.

By the {\it outputs of a contract} we mean the state and gas consumption of transactions, and the emitted events of transactions. To trace the execution of a contract, it would be useful to visualise all of the outputs on a time line. For example if most transactions succeed but an occasional transaction fails, this should stand out in a visualisation. There are many outputs that can in principle be tracked and visualised but the user would soon be overwhelmed. The challenge is to focus on the most useful outputs, and to do so automatically.

Developers of smart contracts for Ethereum often use the Truffle framework to test contracts. The tests are written in JavaScript and the truffle framework ensures that the tests can communicate with the contact when it is deployed on one of the test networks or the main Ethereum network. The Truffle framework requires a directory with number of files as shown below. The directory {\normalsize \verb=contracts=} contains the Solidity code of the contract to be tested (here {\normalsize \verb=Vitaluck.sol=}) as well as a migration contract that is used by the Truffle framework in the deployment phase. The {\normalsize \verb=test=} directory contains the test script for the contract (here {\normalsize \verb=vitaluck.js=}). Section~\ref{subsec:test} gives an example of a test script. We refer the reader to the documentation of the Truffle framework for further details. 

{\normalsize
\begin{verbatim}
Vitaluck/
|-- contracts/
|   |-- Migrations.sol
|   +-- Vitaluck.sol
|-- migrations/
|   +-- 1_initial_migration.js
+-- test/
    |-- vitaluck.js
    +-- truffle-config.js
\end{verbatim}
}

\subsection{A model of replay accuracy}
\label{subsec:model}
We present a model of the differences between the historic execution of a smart contract as recorded on the blockchain and the isolated replay on a minimal testing environment. We discuss the issues that will arise, and propose an experiment to measure their effect on replay accuracy.

To assess the extent to which a contract has been replayed accurately we need a definition of replay accuracy. A strict definition might require that all outputs of each transaction are the same in the replay as in the historic transaction. This would not be a particularly useful definition as there are too many environmental parameters over which we have no control. For example the amount of gas used by a transaction is not necessarily the same on all EVM implementations. Therefore, we propose a definition of replay accuracy that takes the most important outputs into account: the transaction status, and the event parameters.

Status accuracy measures the extent to which the historic status of each transaction can be reached when replaying the contract with historic inputs in a minimal testing environment.
\begin{quote}
{\bf status accuracy} If the status of the first $t \leq T$ transactions of the replay run is the same as the status of the first $T$ transactions of the historic run, we have $t/T$ replay accuracy.
\end{quote}
If the status of two transactions disagrees, the state variables of the historic and the replay run may also disagree. Therefore, we stop counting after the first status disagreement.

Event accuracy measures to what extent the transaction events of the replay transaction agree with the transaction events of the historic transactions. For each event emitted by a contract, each parameter of the historic transaction must be the same as the corresponding parameter of the replay. If a replayed transaction fails where the historic transaction succeeds, there will not be any output for the replay, and the comparison yield false. The definition of event accuracy is:
\begin{quote}
{\bf event accuracy} If the event parameters of the first $t \leq T$ transactions of the replay run are the same as the event parameters of the first $T$ transactions of the historic run, we have $t/T$ event accuracy.
\end{quote}

Status and event accuracy are numbers between 0 and 1. We will sometimes refer to just replay accuracy when it does not matter which of the two is meant.

In the following sections we will discuss how the environment in which a contract is executed can affect replay accuracy. Contracts can interact with the environment in two ways. A contract may (a) use special variables to access the environmental parameters, or (b) it may call, or it may be called, by another contract.

\subsection{Special contracts}
Contracts are in principle deterministic. I.e. running the same sequence of transactions twice with the same inputs should lead to the same sequence of outputs. However, if a contract uses {\it special variables}, replaying a sequence of transactions usually leads to a different sequence of outputs. The special variables are {\normalsize \verb=block=}, which gives access to various properties of the current block, {\normalsize \verb=now=}, the time stamp in seconds of the current block, and {\normalsize \verb=blockhash=}, the hash of any of the last 256 blocks. A contract that uses a special variable will be called a {\it special contract}. A contract that does not use a special variable will be called a {\it regular contract}.

Since special contracts do not necessarily generate the same sequence of outputs when a particular sequence of transaction is replayed, it makes sense to track not just the outputs of one sequence of replayed transactions but also the outputs of a number of replay sequences. To visualise the output of these multiple replays, we can conveniently create a table for every individual output, with rows enumerating the transactions and columns enumerating the individual replay runs. The behaviour of the individual outputs can then be visualised with a heat map. For example the amount of the gas used by each transaction is a useful individual output to track, as illustrated in Figure~\ref{fig:Vitaluck_gasUsed}. We will explain the figure in more detail later.

Special contracts are designed to behave differently each time they are replayed, and we should expect the replay accuracy to be reduced. Therefore we hypothesise that:
{\bf Special contracts have lower replay accuracy than regular contracts.}

\subsection{Dependent contracts}
Smart contracts often call methods of other smart contracts. For example Etheroll (See~\url{https://etherscan.io/address/0xddf0d0b9914d530e0b743808249d9af901f1bd01#code}) is a client that calls methods of a server {\normalsize \verb=OraclizeAddressResolver=} (See~\url{https://etherscan.io/address/0x1d3b2638a7cc9f2cb3d298a3da7a90b67e5506ed#code}). This dependency is transparent because the name of the {\it callee} is explicitly mentioned in the source of the {\it caller}. A fragment of the code of Etheroll is shown below, as an illustration of how a typical client obtains access to a server. The modifier {\normalsize \verb=coupon=} contains a call to the method {\normalsize \verb=getAddress()=} of the address resolver contract {\normalsize \verb=OAR=}. The result is the current address of the Oraclize contract, which is stored in the variable {\normalsize \verb=oraclize=}, ready for further use. The bottom of the code fragment below shows that the type of the hard coded address of the Oraclize proxy is cast into the contract interface type {\normalsize \verb=OraclizeAddrResolverI=}. The result of the type cast is assigned to the variable {\normalsize \verb=OAR=}.

{\normalsize
\begin{verbatim}
contract Etheroll {
    ....
}
contract OraclizeAddrResolverI { // Contract interface
    function getAddress() returns (address _addr);
}
contract usingOraclize {
    OraclizeAddrResolverI OAR;
    OraclizeI oraclize;
    modifier coupon(string code){
        oraclize = OraclizeI(OAR.getAddress()); // Type cast to contract
        ...
    }
...
    OAR = // Type cast to proxy contract interface
        OraclizeAddrResolverI(0x1d3B2638a7cC9f2CB3D298A3DA7a90B67E5506ed);
...
}
\end{verbatim}
}

The hard coded address {\normalsize \verb=0x1d3B...=} is the permanent address of the proxy of the Oraclize contract. The code of the proxy contract {\normalsize \verb=OraclizeAddrResolver=} is shown in full below. The constructor stores the address of the account that deployed the proxy in the variable {\normalsize \verb=owner=}. Only the owner can change this by calling the method {\normalsize \verb=changeOwner=}. The owner of the proxy can also change the address of the server by calling {\normalsize \verb=setAddress()=}. A client of the proxy, such as Etheroll, can only call {\normalsize \verb=getAddress()=} to obtain the current address of the server.

{\normalsize
\begin{verbatim}
contract OraclizeAddrResolver {
    address public addr;
    address owner;
    function OraclizeAddrResolver(){
        owner = msg.sender; // Set initial owner
    }
    function changeOwner(address newowner){
        if (msg.sender != owner) throw;
        owner = newowner; // Change owner
    }
    function setAddr(address newaddr){
        if (msg.sender != owner) throw;
        addr = newaddr; // Change server
    }
    function getAddress() returns (address oaddr){
        return addr; // Return the current server
    }
}
\end{verbatim}
}

The proxy acts as gateway to the server contract. Putting a proxy between the client and the server makes it possible to update the server without having to update the clients. This is useful, as contracts cannot be changed; they can only be redeployed. However, having proxies is also opaque, as the client may rely on specific server functionality that can change without the client's knowledge. If we take the servers point of view, then on the one hand it is useful that any client can make use of the server. On the other hand the server has no way of knowing a-priori which clients will be calling its functions. Unfortunately, current blockchain explorers do not show a-posteriori who called a particular server. This is a significant transparency issue.

Replaying a smart contract without its dependencies usually fails. Therefore, we hypothesise that:
{\bf A contract that casts an address type to a contract type will have lower replay accuracy than a contract without such a type cast.}

\subsubsection{Calls from other contracts}
Detecting calls from other contracts on the blockchain is a problem because with the current Ethereum APIs there is no efficient way of locating all {\it caller} contracts that call a method of a specific {\it callee}. Basically one would have to search over a billion transactions on the Ethereum blockchain to expose all calls. Google Big query offers a public data base of all Ethereum transactions (See~\url{https://console.cloud.google.com/bigquery?p=bigquery-public-data&d=ethereum_blockchain}) that makes such a search possible. Since the service is paid, we do not consider it here as a source of data for our replay tooling. Current blockchain explorers do not show callers from other contracts, but they do show all callees. It should therefore not be difficult to show also the callers.

Replaying a smart contract without calls from other contracts may fail. Therefore, we hypothesise that:
{\bf A contract that is called by other contracts will have lower replay accuracy than a contract without such calls.}

\subsubsection{Ad-hoc address encoding}
\label{subsubsec:address_encoding}
Smart contracts operate on the addresses of externally owned accounts and contract accounts. Since we do not possess the private key associated with the externally owned accounts, we have to replace historic addresses that occur in contracts and transactions by addresses for which we do have the private key in our minimal test environment. The basic set of addresses used by a contract consists of the contract address, and the addresses passed via the {\normalsize \verb=from=} and {\normalsize \verb=to=} parameters of the transactions. Transactions may also carry additional parameters of type address, and there may also be hard-coded addresses in the source of the contract. Most contracts identify hard-coded addresses in the source by giving them the {\normalsize \verb=address=} type, but sometimes they are merely typed as integers. Since all valid addresses are exactly 20 bytes, we assume, heuristically, that we are dealing with an address if we encounter a 40 hex digit integer.

Unfortunately, addresses may also be encoded in an ad-hoc manner. For example the contract Ledger (See~\url{https://etherscan.io/address/0xe6a51bd48f93abcd6c1d532112094044971d8d4e#code}) contains the {\normalsize \verb=multiMint=} method shown below. The elements of the array {\normalsize \verb=bits=} contain an encoding of address and amount pairs. The encoding is the concatenation of a 96-bit value with a 160-bit address. Since we use the standard tools to compile contracts and to replay transactions, we can simply extract all decoded addresses from the debug output from {\normalsize \verb=ganache=}, the EVM client (See~\url{https://github.com/trufflesuite/ganache}). A blockchain explorer could in principle use the same strategy to extract decoded addresses.

{\normalsize
\begin{verbatim}
function multiMint( uint[] bits ) {
    for( uint i=0; i<bits.length; i++ ) {
        address a = address( bits[i]>>96 );
        uint amount = bits[i]&((1<<96) - 1);
        mint( a, amount );
    }
}
\end{verbatim}
}

Assuming that a minority of the historic addresses are encoded in an ad-hoc fashion, we think that there will be a relatively small residue of such addresses in the replay scripts. Since these may cause transactions to fail, we hypothesise that:
{\bf A residue of ad-hoc encoded historic addresses has a negligible effect on replay accuracy.}

\subsection{Research questions}
The above considerations lead to the following research questions:

\textit{Q1:} To what extent does Etherscan provide the information necessary to replay contracts with Truffle?

\textit{Q2:} To what extent does the presence of (a) special variables, (b) address-to-contract type casts, (c) internal transactions, and (d) residual historic addresses influence replay accuracy?

\textit{Q3:} How to track and visualise the most interesting outputs (e.g. status, event parameters) of the contracts?

\subsection{A case study in transparency: Vitaluck}
\label{sec:case_study}
To illustrate the utility of replaying a contract with historic inputs we will elaborate on a case study presented in a recent publication~\cite{Chia2018}. Vitaluck (See~\url{https://etherscan.io/address/0xef7c7254c290df3d167182356255cdfd8d3b400b#code}) is a lottery contract with only 27 transactions. This makes it sufficiently small to be used as a case study.

The problem that we will focus on is not new, but we believe that our way of analysing the problem is more intuitive that the usual source code inspection. While there are several proposals for a good source of randomness in smart contracts~\cite{Chatterjee2019}, most developers use special variables like {\normalsize \verb=now=} as a source of randomness. To some extent miners can manipulate special variables, which makes this problematic. Vitaluck also uses {\normalsize \verb=now=} and we will replay all the transactions of the contract with a timestamp that is perfectly controlled to illustrate the problem of using {\normalsize \verb=now=}.

The smart contract code of Vitaluck in essence uses the time stamp of the current block (viz. {\normalsize \verb=(now*2)%1000=}) as a random number. On the public blockchain this should not be much of a problem as blocks are created about every 14 seconds. However, if {\normalsize \verb=now=} increases by 1 each time a transaction occurs, as it does on our minimal testing environment, the random number used by Vitaluck will only increase by two.

\begin{figure}[t]
  \centering
     \begin{minipage}{.45\linewidth}
       \includegraphics[width=\textwidth, trim={10mm 20mm 10mm 20mm}, clip]{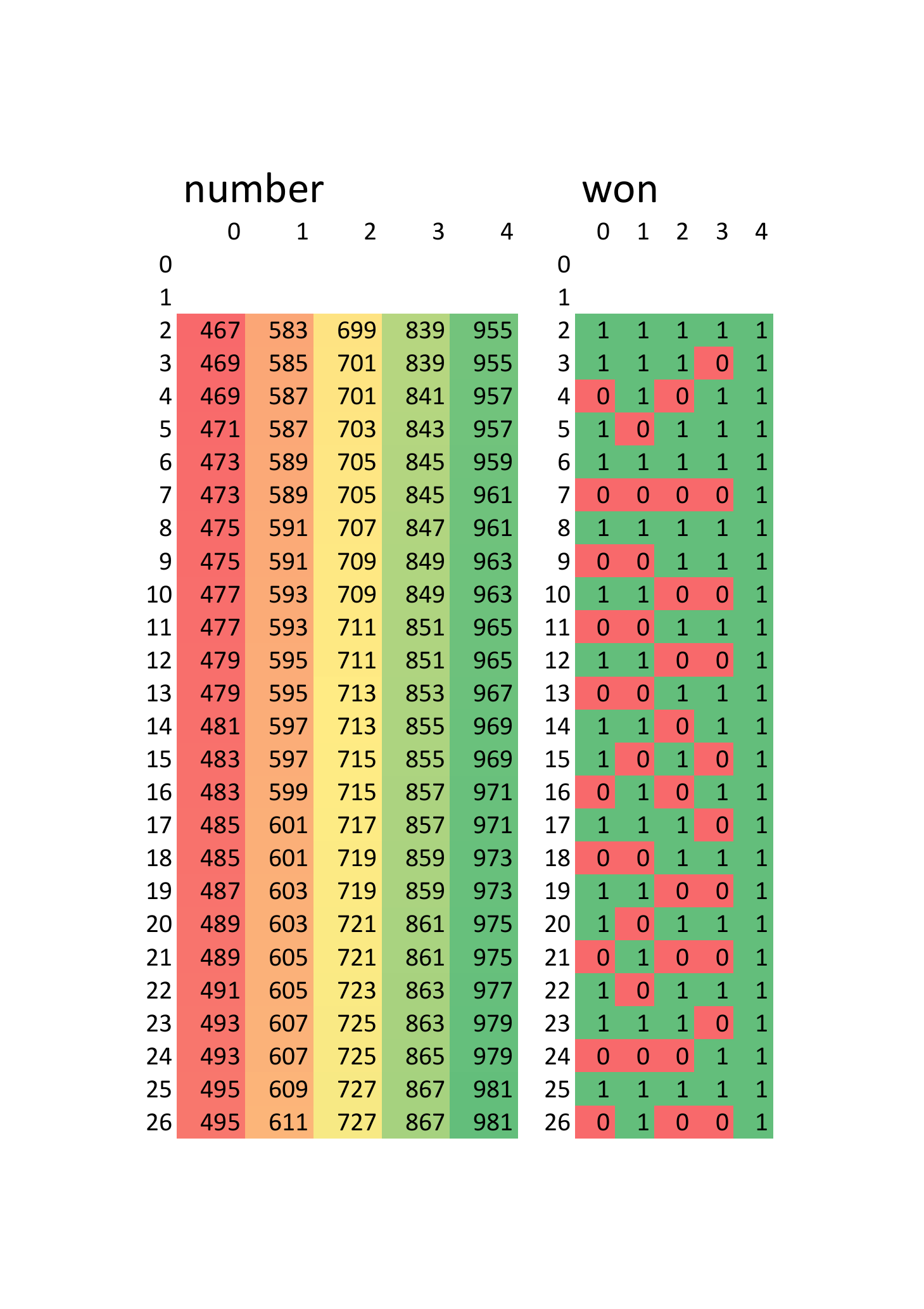}
       \caption{Tracking the outputs of the Vitaluck contract showing how low entropy randomness leads to unfair behaviour.}
       \label{fig:Vitaluck_number_won}
     \end{minipage}
     \hspace{.05\linewidth}
     \begin{minipage}{.45\linewidth}
       \includegraphics[width=\textwidth, trim={10mm 20mm 10mm 20mm}, clip]{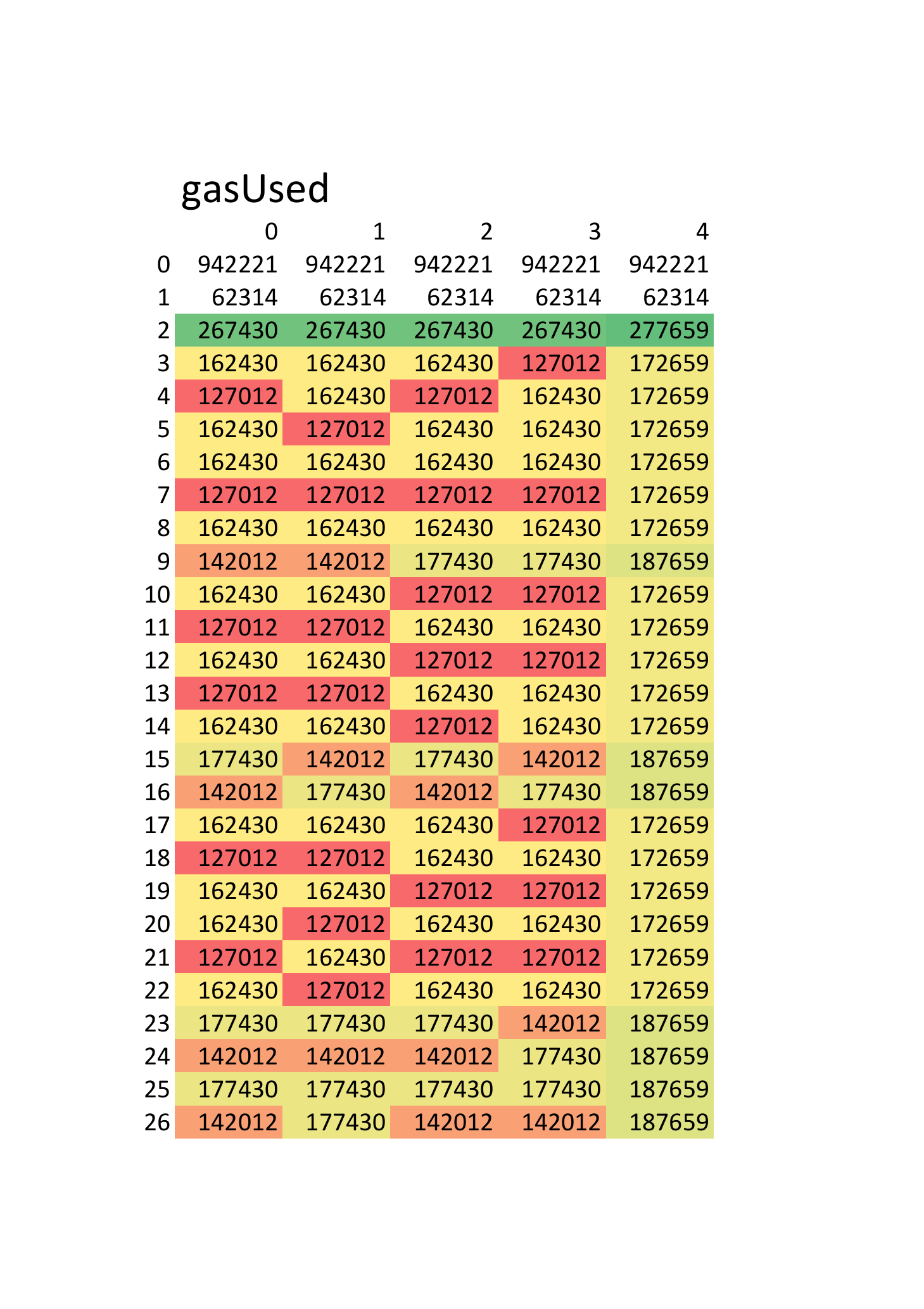}
       \caption{Tracking the gas used by the Vitaluck contract showing how information clearly visible on the blockchain correlates perfectly with the functionality.}
       \label{fig:Vitaluck_gasUsed}
     \end{minipage}
\end{figure}

The heat maps of Figure~\ref{fig:Vitaluck_number_won} illustrate the problem. In each of the two panels, there are 5 columns, labelled 0 to 4. These correspond to 5 replay runs of the sequence of 27 transactions of the contract. Each row corresponds to a transaction. Rows 0 and 1 represent the contract deployment. Rows 2 to 26 correspond to one execution of the {\normalsize \verb=Play=} method. The left panel tracks {\normalsize \verb=number=}, which is one of the parameters of the {\normalsize \verb=NewPlay=} event, that is emitted by the {\normalsize \verb=Play=} transaction. The last column of the left most panel (in green) shows that in run 4, the value of {\normalsize \verb=number=} is always above 900. The source code states, that the player then has a 99.9\% of winning again. This is also illustrated by the value of {\normalsize \verb=won=} for replay run 4 in the last column of (also in green) of the right most panel. This shows that each transaction has produced a winner. The tables for {\normalsize \verb=number=}, and {\normalsize \verb=won=} show that a heat map is sufficient to show the problematic behaviour of what was intended to be a fair lottery. The heat maps are produced automatically and they can only be produced because we can replay the historic transactions with different block times.

Vitaluck also leaks information through its gas usage. Figure~\ref{fig:Vitaluck_gasUsed} shows the amount of gas used for each of the 27 transactions of the five replay runs. The colours in the heat map illustrate the correlation with the rightmost panel in Figure~\ref{fig:Vitaluck_number_won}. Inspecting the code confirms that if there is a winner, the contract executes more code and therefore consumes more gas. The gas usage is recorded in the blockchain; hence there should not be any expectations of confidentiality. In the case of Vitaluck the fact that there has been a winner is completely transparent.

\begin{figure}[t]
  \centering
    \includegraphics[width=0.47\textwidth, trim={0mm 0mm 0mm 0mm}, clip]{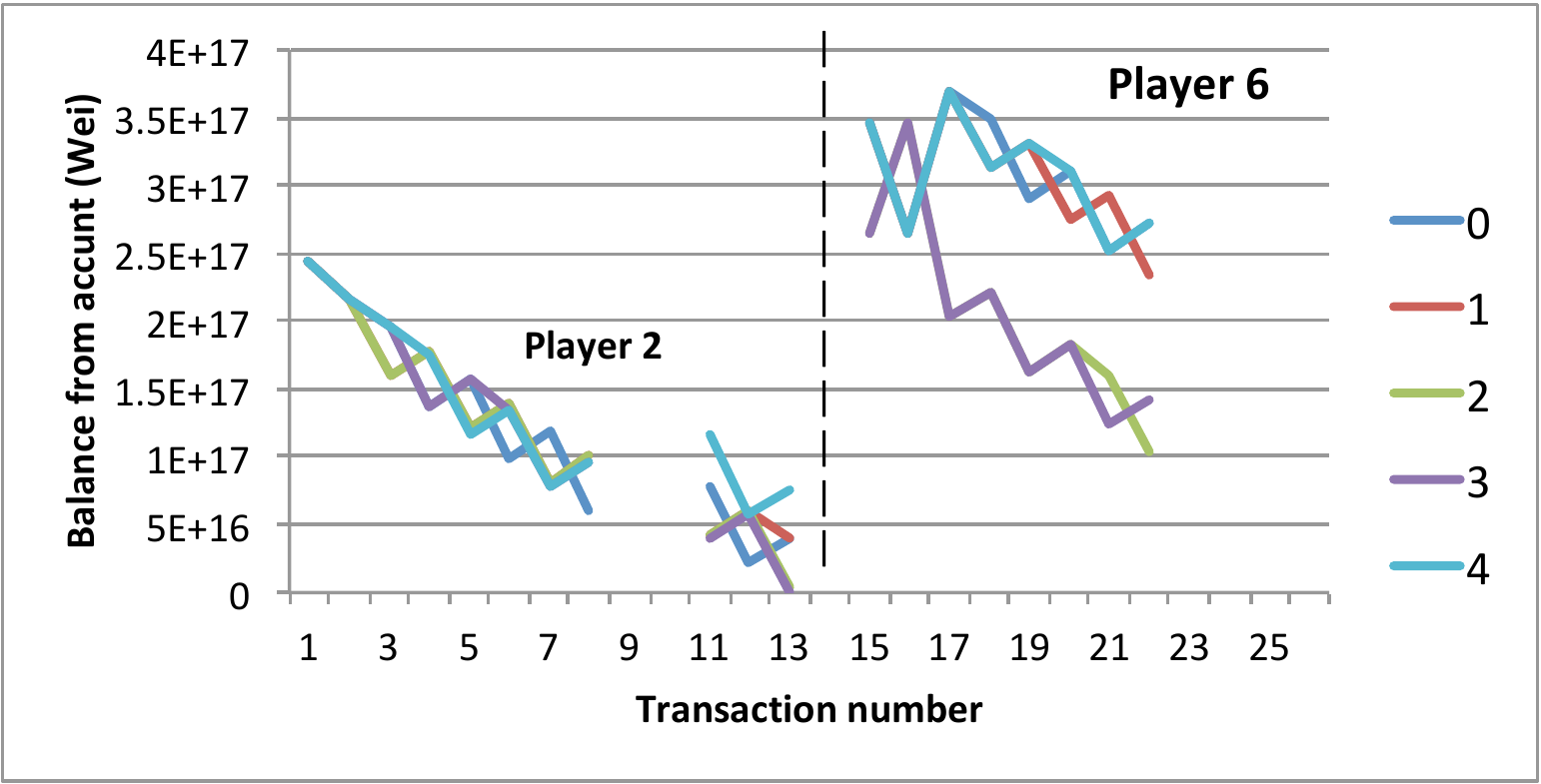}
      \caption{Tracking the balance of the accounts of Player 2 (during transactions 1-13) and Player 6 (during transactions 15-25) calling the Play method, showing that playing more often reduces the balance by a small amount that that it increases the balance by winnings.}
    \label{fig:Vitaluck_fromBalance}
\end{figure}

A lottery contract would not be successful if the users did not on occasion win. We can automatically track the balance of relevant accounts to investigate this. For example, Figure~\ref{fig:Vitaluck_fromBalance} tracks the balance of different users executing the {\normalsize \verb=Play=} method of the contract. The horizontal axis gives the transaction number and the vertical axis the balance (in excess of the minimum balance of all accounts) of the relevant player. The colours in the graph correspond to replay runs. Each of the 5 different graphs gives the output of one of the 5 replay runs. The general trend of the balance of a player is to go down, unless he/she wins, when it goes up significantly. During transactions 1-10 and 13-14 only player 2 is active. Player 6 is only active during transactions 17-24. Other players are active during other transactions but these have been removed from the graph to avoid clutter. The graph shows that Player 6 has won once in two replay runs (2 and 3), at transaction 17, where his balance increased considerably, and continues after his win. Player 2 has abandoned the game at transaction 13. All graphs shown in the case study are easily produced from the output of the replay runs. We believe that the case study illustrates that being able to replay transactions with historic inputs, while at the same time varying the block time provides insights in the functioning of a smart contract. This increases transparency.

\section{Method}
\label{sec:method}
We propose the following method to answer the research questions.
\begin{itemize}
\item Build a tool {\it ContractVis} that (a) downloads a verified smart contract and its historic transactions from Etherscan, (b) generates a {\normalsize \verb=truffle test=} script that replays the transactions, and (c) generates output during the replay that can be analysed and visualised.
\item Draw a random sample of $N$ verified smart contracts. Since every failing replay has to be inspected manually to classify the reasons for failure, it is not feasible to process more than a sample.
\item For each of the randomly sampled smart contracts, replay the first $T$ transactions of the contract, and perform each replay run $R$ times. We limit the number of transactions for two reasons. Firstly, the longer the replay run, the more likely it is that a replay will fail. Therefore, making all replay runs the same length allows for a fair comparison between contracts. Secondly, the load on Etherscan of downloading large numbers of transactions is significant.
\item Analyse the outputs of each of the $R$ replay runs, and compute the status and event replay accuracy.
\item Analyse the differences and similarities of the selection of contracts, specifically looking for reasons why transactions fail.
\end{itemize}
In Section~\ref{sec:results} we will discuss what values we have chosen for $N$, $R$, and $T$.

\subsection{Generating replay scripts}
\label{subsec:generating}
For each transaction, starting with the create transaction, ContractVis generates a test in JavaScript that can be replayed by the {\normalsize \verb=truffle test=} command. The automatically generated test takes the following actions:
\begin{itemize}
\item Deploy the contract with the historic instantiation parameters as found on the blockchain, except that historic addresses are replaced.
\item Issue all $T$ historic transactions (modulo address translation). The status of the historic and the replayed transaction are compared to calculate the status accuracy.
\item If a historic transaction has failed, we force the replayed transaction also to fail, by passing an illegal {\normalsize \verb=from=} address to the transaction. Otherwise the state of a replayed contract would no longer be consistent with historic state.
\item Get the balance of the {\normalsize \verb=from=} and {\normalsize \verb=to=} accounts used in all $T$ transactions on the blockchain, and generate calls to obtain the balances via the appropriate API calls. This can be used to compare the effect of the transaction on the balances of the accounts in the replay with the historic balances.
\item Extract all historic event parameters emitted by all $T$ transactions on the blockchain, and generate API calls to extract the events also on the test environment. This is used to compare the event parameters of the replay run to the historic parameters, and to calculate the event accuracy.
\item After all $T$ transactions, generate calls to each pure function of the contract with random arguments. This is intended to simulate actions that the DApp could have performed, and thus to reduce the opacity of the contract. \end{itemize}

ContractVis downloads the source of the contract, the application binary interface (ABI), the first $T$ transactions, the event logs, and the balances of the addresses involved after each transaction from Etherscan. The tool pauses regularly for a few seconds to comply with the rate limiting policies of Etherscan.

The central data type is the {\it Prototype} format, which is a generalisation of the decoded ABI format. The Etherscan API provides a variety of information in different formats, all of which is converted into prototypes. The code generation converts prototypes into JavaScript code.

The appropriate {\it emscripten} version of the solidity compiler (See~\url{https://github.com/ethereum/solc-bin/blob/gh-pages/bin/list.json}) is used to generate the abstract syntax tree from the source, which is traversed to collect the hard coded addresses from the contracts. These historic addresses as well as the {\normalsize \verb=from=} and {\normalsize \verb=to=} addresses used by the transactions are replaced by addresses generated by {\normalsize \verb=ganache=}. These addresses are generously stocked with Ether. The maximum number of addresses that can be used in the source code and the transactions together has been limited to 1000 but this can be changed easily. The list of transactions will be truncated if more addresses than the limit occur in the source code and the transactions.

Some verified contracts are created in an internal transaction. Etherscan provides the hash of this transaction, so that we can access the internal transactions via the parity trace. For example the contract {\normalsize \verb=Congress=} (See~\url{https://etherscan.io/address/0x97282a7a15f9beadc854e8793aae43b089f14b4e#code}) is created by this transaction~\url{https://etherscan.io/vmtrace?txhash=0x0233ee95d6b34e3d67b380d0c7d58dca533ce5e3a07ed0b64c3ef3aa60aedbe6&type=parity}.

Once all data has been collected, the relevant Truffle contract, migration, and test code is generated. The migration script deploys the contract with its constructor arguments. The test script encodes the $T$ transactions with their parameters in a {\normalsize \verb=truffle test=}. The tests generate a variety of useful data in JSON format for further processing.

\subsection{Example: Vitaluck test script}
\label{subsec:test}
The JavaScript code below shows the essence of the test script generated for the Vitaluck case study. To avoid clutter, we have removed all boilerplate from the code. The array {\normalsize \verb=adList=} contains the addresses that are used during the replay instead of the historic addresses. These replay addresses are generated by {\normalsize \verb=ganache=} and provided unlocked as the {\normalsize \verb=accounts=} parameter to the migration script. For example the {\normalsize \verb=from=} address in the create transaction is {\normalsize \verb=adList[5]=}, because the corresponding historic transactions used the corresponding entry in the historic address list. So the {\normalsize \verb=adList=} array is the implementation of the address translation from historic blockchain addresses to addresses of the minimal test environment.

To ensure that all transactions happen at the same time as the historic transactions, the test does not use instamining but mines each block explicitly, with the time stamp of the historic transaction.

{\normalsize
\begin{verbatim}
const Vitaluck = artifacts.require( "./Vitaluck.sol" ) ;
var adList = null ; // List of addresses
var deployed = null ; // The deployed contract
contract( "Vitaluck", function( accounts ) { 
    it( "Constructor(  )", async function( ) { // Deploy the contract
        await support.minerStop( ) ; // Stop instamining
        adList = [    "0x0000000000000000000000000000000000000000",
                      "0x0000000000000000000000000000000000000001",
                      "address(this)" 
                 ].concat( accounts ) ) ;
        const txOptions = // Create transaction
            {from: adList[5], to: 0, value: "0"} ;
        const txCall = {inputs: [], name: "Vitaluck", outputs: [], type: "function"} ;
        const txRequest = // Deploy the contract and call the constructor
            Vitaluck.new( txOptions ) ; 
        await new Promise( (resolve) => {txRequest.on( "transactionHash", resolve )} ) ;
        await support.mineBlockWithTimestamp( 0, 1516973359 ) ;
        deployed = await txRequest ; 
        const txReceipt = // Get the transaction details
                await web3.eth.getTransactionReceipt( deployed.transactionHash ) ;
        deployed.address = txReceipt.contractAddress ;
        adList[2] = txReceipt.contractAddress ;
    } ) ;
    
    it( "Play(  )", async function( ) { // The first test
        const txOptions = {from: adList[3], to: adList[2], value: "50000000000000000"}
        const txCall = // Regular transaction
            {inputs: [], name: "Play", outputs: [], type: "function"} ;
        const txRequest =  // Call the Play method
            deployed.methods[ "Play()" ]( txOptions ) ;
        await new Promise( (resolve) => {txRequest.on( "transactionHash", resolve )} ) ;
        await support.mineBlockWithTimestamp( 1, 1516973566 ) ; 
        const txResult = await txRequest ;
        const fromBalance = // Get the balance
            await web3.eth.getBalance( adList[3], txResult.blockNumber ) ) ;
    } ) ;
    ...
} )
\end{verbatim}
}

The test script contains many {\normalsize \verb=it=} blocks of which we show only the first two. The first {\normalsize \verb=it=} block ensures that the contract has been deployed, and that the {\normalsize \verb=adList=} array is initialised.

The first {\normalsize \verb=it=} block sends {\normalsize \verb=value: "0"=} to address {\normalsize \verb=to:0=}, which indicates a create transaction. The cost of creating the transaction is paid for by the account with address {\normalsize \verb=from: adList[5]=}. The {\normalsize \verb=it=} block completes once the create transaction has been mined.

The second {\normalsize \verb=it=} block fires the first transaction, which calls the {\normalsize \verb=Play=} method of the Vitaluck contract. {\normalsize \verb=adList[1]=} is the account that created the contract. So {\normalsize \verb=txOptions=} specifies that $50\times10^{15}$ Wei is sent from account {\normalsize \verb=adList[2]=} to the contract address {\normalsize \verb=adList[1]=}. Once the transaction has been mined, the balance of the two accounts involved in the transaction is retrieved with the calls to {\normalsize \verb=web3.eth.getBalance(.)=}. All values can be used for further analysis and visualisation.

\subsection{Example: Vitaluck getter script}
Ethereum does not have programmatic reflection capabilities, and therefore the internal state of a contract is transparent only to the contract itself. Simply replaying an existing transaction will not expose internal state beyond what the contract event parameters and existing getters expose. However, if important getters are missing, they can only be added if the bytecode of the contract is changed. This is not possible for an existing deployment on the public blockchain. Since our replay facility does re-deploy a contract, it is easy to add any missing getters to the contract. The program fragment below shows sample code generated for the Vitaluck contract that calls existing getters on the state. Vitaluck provides sufficient getters already but it would be easy to add any missing getters.

{\normalsize
\begin{verbatim}
async function getters( deployed ) {
    console.log( await deployed.GetStats() );
    console.log( await deployed.GetLastBetUser(...) ) ;
    console.log( await deployed.GetWinningAddress() );
    console.log( await deployed.ownerBetsCount(...) ) ;
    console.log( await deployed.GetUserBets(...) ) ;
    console.log( await deployed.GetBet(...) ) ;
    console.log( await deployed.GetCurrentNumbers() ) ;
}
\end{verbatim}
}
If a getter has parameters, these are randomly chosen by ContractVis. To avoid clutter, in the code above we have replaced the random parameters by three dots.

\section{Results}
\label{sec:results}
On 1 January 2019 there were 54,179 verified contracts on Etherscan. From these we selected the contracts that were not too old, and which had at least $T=50$ transactions. This lower limit was set to avoid contracts with too few transactions to generate replay scripts from. We excluded older contracts that had to be compiled by Solidity compilers before version 0.4.9 because of compatibility issues with those old compilers. From the list of 11,234 remaining contracts we selected approximately 10\% using uniform sampling ($N=1120$). This random sample of contracts was downloaded, processed by ContractVis, compiled, deployed, and replayed with the Truffle tools.

\begin{center}
\begin{table}[ht]
\caption{Descriptive statistics of the verified smart contracts sampled from Etherscan}
\centering
\begin{tabular*}{0.73\textwidth}{l *{5}{r} }
\toprule
            &    SLOC & Transactions &     Ether &  Instruction space
                                                            & Coverage \\
            &$N=1111$ &     $N=1120$ &  $N=1120$ & $N=1037$ & $N=1035$ \\
\midrule
Minimum     &     9.0 &         51.0 &       0.0 &     43.0 &    6.1\% \\
Maximum     & 5,039.0 &  5,553,551.0 & 184,419.3 &  14162.0 &   98.4\% \\
Average     &   417.5 &     13,732.5 &     249.8 &   3474.9 &   51.4\% \\
Stdev       &   403.0 &    174,321.3 &   5,736.0 &   2159.5 &   16.2\% \\
\bottomrule
\end{tabular*}
\label{table:descriptives}
\end{table}
\end{center}

\begin{center}
\begin{table}[ht]
\caption{Compiler versions}
\centering
\begin{tabular*}{0.37\textwidth}{l r r }
\toprule
Year    & compiler& registration \\
\midrule
2017    &   663   &  657 \\
2018    &   457   &  463 \\
Total   &  1120   & 1120 \\
\bottomrule
\end{tabular*}
\label{table:compiler}
\end{table}
\end{center}

\subsection{Descriptive statistics}
Table~\ref{table:descriptives} presents statistics that characterise our sample of verified smart contracts.  The average Source Lines of Code (SLoC) statistic of our contracts is 417.5. The shortest contract has 9 SLoC (See {\normalsize \verb=Pgp=}~\url{https://etherscan.io/address/0xa6a52efd0e0387756bc0ef10a34dd723ac408a30#code}) and the largest has 5,039 SLoC (See {\normalsize \verb=DemocracyKit=}~\url{https://etherscan.io/address/0x705cd9a00b87bb019a87beeb9a50334219ac4444#code}).  Heged\"us~\cite{Hegedus2019} reports an average size of 57.7 with a standard deviation of 106.4 per contract for a collection of 40,352 source code files from Etherscan. Our SLOC statistic is for source code files, which, according to Heged\"us, contain 4.44 contracts on average. Taking this into account reduces our average to $419.5 / 4.44 = 94.0$, which, while larger than 57.7, is still within one standard deviation from the Heged\"us average.

On 1 Jan 2019, the average number of transactions executed against our sample was 13,732.5, and the average balance of the contracts was 249.8 Ether. Both of these statistics exhibit the typical long tail distribution, with few contracts contributing most to both parameters, and a long tail of contracts that contribute little.

The last two columns of Table~\ref{table:descriptives} shows descriptives of bytecode generated for each contract. The instruction space gives the number of EVM instructions. The coverage indicates the percentage of EVM instructions that are executed by the first 50 historic transactions. Both statistics vary in roughly he same way as the SLOC statistic.

Table~\ref{table:compiler} shows a breakdown by year of the number of contracts. The compiler column shows the number of contracts compiled by Solidity versions from 2017 (0.4.9 to 0.4.19) and from 2018 (0.4.20 and up). The registration column shows the number of contracts that were registered on Etherscan in a particular year. The distribution over the two years is more or less even.

\subsection{Failed versus runnable contracts}
A total of 83 contracts out of $N=1120$, i.e. a mere 7.4\% could not be compiled and/or deployed. Most of the issues are inherent to the limitation of a minimal testing environment. The issues are:
\begin{itemize}
\item 60 contracts failed during the deployment phase because of missing dependencies.
\item 17 smart contracts depend on external library contracts, and therefore failed during deployment. 
\item For four contracts, Etherscan does not provide the correct arguments to a transaction. This is probably due to issues with the Etherscan database.
\item Two contracts generated compiler errors. One caused a heap overflow and the other a stack overflow. This is surprising, as we have used the same compiler version and optimisation settings as during the contract verification by Etherscan.
\end{itemize}

The remaining $1120-83=1037$ contracts will henceforth be called the {\it runnable} contracts. For each runnable contract, we generated a {\normalsize \verb=truffle test=} script from the first $T=50$ transactions as recorded on the Ethereum blockchain.

\begin{center}
\begin{table}[ht]
\caption{Cross tables of status and event accuracy against dependencies on the environment.}
\centering
\begin{tabular*}{0.87\textwidth}{l l r r r r | r r r l}
\toprule
           &          & failed  & intermediate  & perfect & total   & failed & intermediate & perfect & total \\
\midrule
          & \multicolumn{9}{l}{1. Contract uses special variables ($N=1120$, $\chi^2=46.2$, $p<0.001$)} \\
\midrule
Status    & no        & 30      & 150   & 420   & 600     &  5.0\%  & 25.0\% & 70.0\%       & 100.0\% \\
accuracy  & yes       & 53      & 205   & 262   & 520     & 10.2\%  & 39.4\% & 50.4\%       & 100.0\% \\
          & Total     & 83      & 355   & 682   & 1120    &  7.4\%  & 31.7\% & 60.9\%       & 100.0\% \\
\midrule
          & \multicolumn{9}{l}{2. Contract uses special variables ($N=1120$, $\chi^2=45.5$, $p<0.001$)} \\
\midrule
Event     & no        & 30      & 139   & 431   & 600     &  5.0\%  & 23.2\% & 71.8\%       & 100.0\% \\
accuracy  & yes       & 53      & 189   & 278   & 520     & 10.2\%  & 36.3\% & 53.5\%       & 100.0\% \\
          & Total     & 83      & 328   & 709   & 1120    &  7.4\%  & 29.3\% & 63.3\%       & 100.0\% \\
\midrule
          & \multicolumn{9}{l}{3. Contract uses type casts from address to contract ($N=1120$, $\chi^2=143.0$, $p<0.001$)} \\
\midrule
Status    & no        & 18      & 90    & 417   & 525     &  3.4\%  & 17.1\% & 79.4\%       & 100.0\% \\
accuracy  & yes       & 65      & 265   & 265   & 595     & 10.9\%  & 44.5\% & 44.5\%       & 100.0\% \\
          & Total     & 83      & 355   & 682   & 1120    &  7.4\%  & 31.7\% & 60.9\%       & 100.0\% \\
\midrule
          & \multicolumn{9}{l}{4. Contract uses type casts from address to contract ($N=1120$, $\chi^2=73.8$, $p<0.001$)} \\
\midrule
Event     & no        & 18      & 107   & 400   & 525     &  3.4\%  & 20.4\% & 76.2\%       & 100.0\% \\
accuracy  & yes       & 65      & 221   & 309   & 595     & 10.9\%  & 37.1\% & 51.9\%       & 100.0\% \\
          & Total     & 83      & 328   & 709   & 1120    &  7.4\%  & 29.3\% & 63.3\%       & 100.0\% \\
\midrule
          & \multicolumn{9}{l}{5. Contract is called by other contracts ($N=1120$, $\chi^2=44.1$, $p<0.001$)} \\
\midrule
Status    & no        & 41      & 174   & 470   & 685     & 6.0\%   & 25.4\% & 68.6\%       & 100.0\% \\
accuracy  & yes       & 42      & 181   & 212   & 435     & 9.7\%   & 41.6\% & 48.7\%       & 100.0\% \\
          & Total     & 83      & 355   & 682   & 1120    & 7.4\%   & 31.7\% & 60.9\%       & 100.0\% \\
\midrule
          & \multicolumn{9}{l}{6. Contract is called by other contracts ($N=1120$, $\chi^2=107.2$, $p<0.001$)} \\
\midrule
Event     & no        & 41      & 130   & 514   & 685     &  6.0\%  & 19.0\% & 75.0\%       & 100.0\% \\
accuracy  & yes       & 42      & 198   & 195   & 435     &  9.7\%  & 45.5\% & 44.8\%       & 100.0\% \\
          & Total     & 83      & 328   & 709   & 1120    &  7.4\%  & 29.3\% & 63.3\%       & 100.0\% \\
\midrule
          & \multicolumn{9}{l}{7. Contract contains ad-hoc encoded historic addresses ($N=1120$, $\chi^2=9.4$, $p=0.009$)} \\
\midrule
Status    & no        & 67      & 319   & 623   & 1009   &  6.6\%   & 31.6\% & 61.7\%       & 100.0\% \\
accuracy  & yes       & 16      &  36   &  59   &  111   & 14.4\%   & 32.4\% & 53.2\%       & 100.0\% \\
          & Total     & 83      & 355   & 682   & 1120   &  7.4\%   & 31.7\% & 60.9\%       & 100.0\% \\
\midrule
          & \multicolumn{9}{l}{8. Contract contains ad-hoc encoded historic addresses ($N=1120$, $\chi^2=9.2$, $p=0.01$)} \\
\midrule
Event     & no        & 67      & 295   & 647   & 1009   &  6.6\%   & 29.2\% & 64.1\%       & 100.0\% \\
accuracy  & yes       & 16      &  33   &  62   &  111   & 14.4\%   & 29.7\% & 55.9\%       & 100.0\% \\
          & Total     & 83      & 328   & 709   & 1120   &  7.4\%   & 29.3\% & 63.3\%       & 100.0\% \\
\bottomrule
\end{tabular*}
\label{table:cross_table}
\end{table}
\end{center}

\subsection{Replay agreement}
To assess the extent to which replay runs agree with historic runs we operationalise the model of \ref{subsec:model} by defining four independent variables and two dependent variables. The independent variables are:
\begin{itemize}
\item Whether a contract uses special variables (no, yes);
\item Whether a contract uses type casts from address to contract (no, yes);
\item Whether a contract is called by other contracts (no, yes);
\item Whether a contract contains ad-hoc encoded historic addresses (no, yes).
\end{itemize}
The dependent variables are:
\begin{itemize}
\item Whether the transaction status of replaying is (a) {\it failed} because the contract could not be deployed, (b) {\it intermediate} because at least one replayed transaction terminated with a different status than the historic transaction (c) {\it perfect} because all transactions agree with the historic transaction on the termination status.
\item Whether the event output is (a) failed because the contract could not be deployed, (b) intermediate because at least one replayed transaction failed to emit an event or disagreed with the historic event on a parameter (c) perfect because all transactions agree with the historic events on all parameters.
\end{itemize}
To assess the effect of the independent variables on the dependent variables we calculate $4\times2=8$ cross tables.
 Cross table 1 of Table~\ref{table:cross_table} shows that 600 (53.6\%) of the runnable contracts are regular (i.e. Uses special variables=no). For the majority (420) of those regular contracts, the replay runs are perfectly status accurate. For the 520 (46.4\%) special contracts, the number of perfectly status accurate contracts (262) is much closer to the number of intermediate status contracts (205). The difference between special and regular contracts is statistically significant ($p < 0.001$). The same conclusion holds for event accuracy in cross table 2 of the table.

Cross table 3 shows that of the 525 (46.9\%) contracts without type casts from address to contract, the majority (417) can be replayed with perfect status accuracy. Of the 595 (53.1\%) contracts that do contain type casts, coincidetally the same number can be replayed with perfect status accuracy (265) as with imperfect status accuracy (265). This difference between contracts with and without type casts is also statistically significant. The analysis of the contracts with respect to the event accuracy, in cross table 4, gives similar, significant results.

Cross table 5 shows that of the 685 (61.1\%) contracts without callers 470 can be replayed perfectly. If there are callers, the difference between perfect and intermediate accuracy disappears. The same observations apply to the event accuracy statistics (See cross table 6). The results are significant, 

Finally, cross tables 7 and 8 show that 111 (9.9\%) contracts contains a residue of ad-hoc encoded historic addresses. Since this is a relatively small fraction of the contracts, the p-values tend to be higher, but the difference is still significant.

We conclude that the presence of special variables and dependencies has a large and significant effect on replay accuracy. The presence of ad-hoc encodings of historic addresses has a smaller, but still a statistically significant effect on the replay accuracy. Given that in each of the 8 cross tables the difference between intermediate and perfect is larger in the "no" rows than in the "yes" rows is evidence that all the dependent variables affect the status accuracy in a statistically significant manner.

\begin{center}
\begin{table}[ht]
\caption{A breakdown of the historic-replay transaction pairs comparing to what extent replayed transaction terminated in the same status as the historic transactions.}
\centering
\begin{tabular*}{0.87\textwidth}{l l *{8}{r} }
\toprule
             &            & \multicolumn{8}{c}{Status of the replayed trasaction, ($N=1120$, $\chi^2=46,033$, $p<0.001$)} \\
             &            & failed    & out of gas& success   & Total & failed    & out of gas& success  & Total  \\
\midrule
status of    &  failed    & {\bf 2193}& 49        & 51        & 2293  & 95.6\%    &  2.1\%    &  2.2\%   & 100.0\% \\
the original &  out of gas& 4         & {\bf 3735}& 0         & 3739  &  0.1\%    & 99.9\%    &  0.0\%   & 100.0\% \\
transaction  &  success   & 9068      & 1024      &{\bf 34806}& 44898 & 20.2\%    &  2.3\%    & 77.5\%   & 100.0\% \\
\midrule
             &  Total     & 11265     & 4808      & 34857     & 50930 & 22.1\%    &  9.4\%    & 68.4\%   & 100.0\% \\
\bottomrule
\end{tabular*}
\label{table:transactions}
\end{table}
\end{center}

The 1120 contracts generated 55,930 historic transactions, which is on average 49.9 transactions per contract. This is slightly less than 50, since the number of transaction on some contracts was limited do to excessive use of historic addresses (See Section~\ref{subsec:generating}).

Table~\ref{table:transactions} shows a cross table of the historic-replay transaction pairs. The transactions of most pairs terminate in the same state (indicated in bold face), in total 40,734 (72.8\%) of perfect agreement. Most of the 9,068 transactions that succeeded in the historic run but failed in the replay did so because of missing dependencies. A few transactions (51) succeed where the historic transaction failed, usually due to the generous Ether balance in the replay.

For 0.8\% of the transactions, the reason for failure is specified by the code. This is usually a mismatch between the environment-related parameters and the expectations of the contracts. Unfortunately, not many contracts follow this best practice of specifying a reason for failure. 

\begin{figure}[ht]
  \centering
    \includegraphics[width=0.47\textwidth, trim={0mm 0mm 0mm 0mm}, clip]{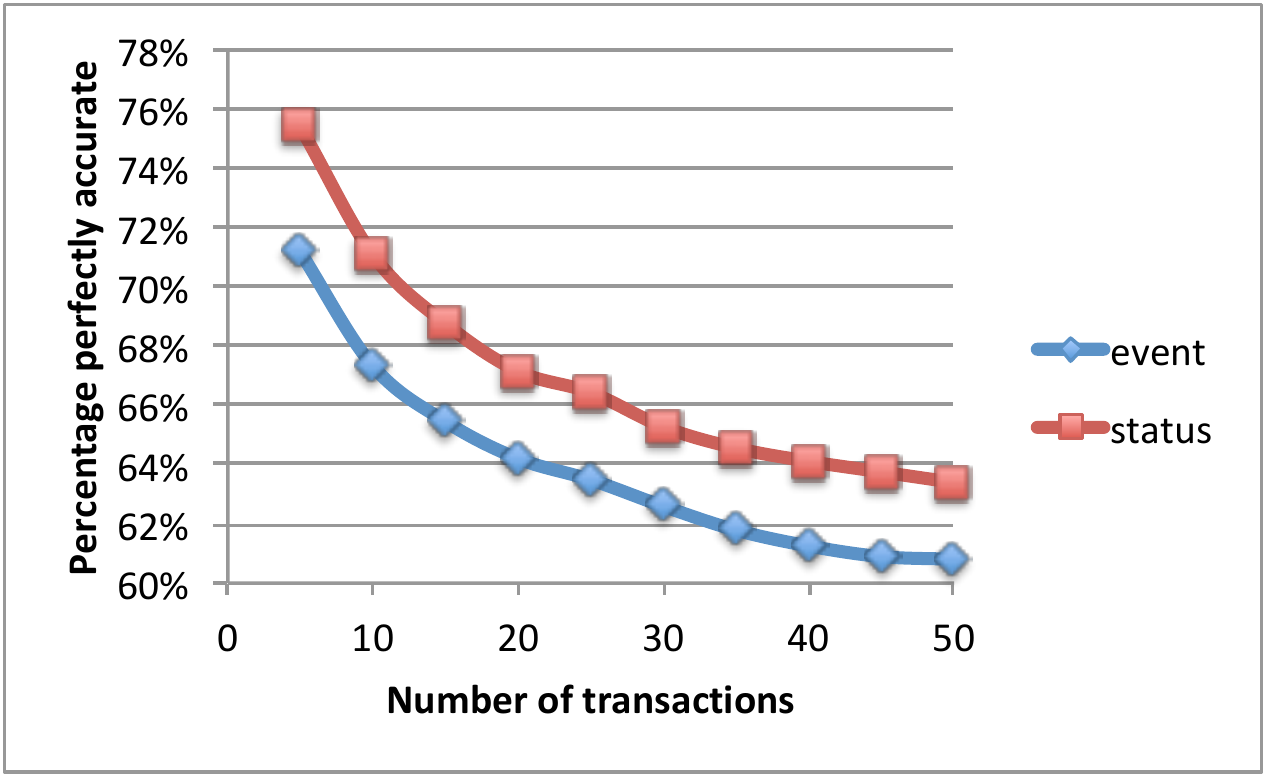}
      \caption{Senstivity analysis of the replay accuracy as a function of the number of transactions in the replay scripts.}
    \label{fig:sensitivity_nTx}
\end{figure}

\subsection{Sensitivity Analysis to the value of $T$}
We have investigated to what extent replay accuracy is sensitive to the choice of $T$, which determines how many transactions are replayed. The lower the value of $T$, the better the change of achieving perfect accuracy, but at the same time the less useful the test script is.  Figure~\ref{fig:sensitivity_nTx} shows that there is less than 15 per cent points difference in accuracy for both the status and the event accuracy between low and high values of $T$. This indicates that the accuracy results are only marginally sensitive to the value of $T$. Values of $T$ below 10 should probably be avoided, as the test scripts would become rather short.

\section{Recommendations}
\label{sec:recommendations}
We provide five recommendations for blockchain explorers that would increase the transparency of smart contracts on Ethereum. We believe that all recommendations can be implemented using the data that blockchain explorers already collect, and some program analysis techniques.

\subsection{Indicate environmental dependencies}
About half the contracts of our sample depend on environmental parameters, such as the time stamp of the current block. A blockchain explorer should be able to list such a dependency. For example knowing that a certain method can only be called during a certain time period increases transparency. This basic information can be used in automatic contract classification, which would increase the transparency of contracts further. An example is automatic token classification~\cite{Froewis2019}.

\subsection{Decode ad-hoc encodings of addresses}
Smart contracts are free to encode addresses in an ad-hoc manner. Such encodings make it more difficult for a blockchain explorer to track the addresses involved, thus reducing transparency. In Section~\ref{subsubsec:address_encoding} we discussed the {\normalsize \verb=multiMint=} method, which ingests about 25 ad-hoc encoded addresses per transaction. This means that just a few transactions can hide large numbers of addresses from the blockchain explorer. Our replay tooling determines automatically whether ad-hoc encodings are used. In total we found 11,485 addresses in the first 50 transactions sent to the 110 contracts with residual historic addresses, hence the problem is not restricted to just a few contracts or a few transactions. We believe that being able to decode ad-hoc encodings involving addresses would be a useful addition to existing blockchain explorers that can have a significant impact. It should be possible to implement this with the appropriate program analysis techniques.

\subsection{Provide detailed reasons for transaction failure}
Out of 50930 historic transactions, 2,193 failed (5.4\%). If the reason for failure is clear, and readily communicated to the user, this contributes to transparency. Consider as an example the game Wallie (See~\url{https://etherscan.io/address/0xe5af8907776fd5f1bb069369fd398ad33102751e#code}). Most methods of the contract check whether the game has started, and issue an appropriate message when this is not the case. The modifier shown below implements this. The {\normalsize \verb=require=} predicate has a second parameter, which tells the initiator of the transaction that the game has not been started when {\normalsize \verb|is_started = false|}.

{\normalsize
\begin{verbatim}
// Check the start of the project
modifier isStarted() {
    require( is_started == true, "Project not started" );
    _;
}
\end{verbatim}
}

The first historic transaction of the Wallie contract was a test to make sure that the message is issued. (See~\url{https://etherscan.io/tx/0x1616021c30a50c2d066b481b6a89f90dce072ace799a6bd1b0e5af7d8e65d02d}). Indeed the transaction has failed, but the message ``Project not started'' is not shown by Etherscan. The message is also not shown by the {\normalsize \verb=geth=}, {\normalsize \verb=parity=} and {\normalsize \verb=remix=} debug traces, nor by any other blockchain explorer. For all 2,193 historic transactions in our sample that failed, not a single meaningful message could be found. The developer has to be content with only three different messages: {\it Reverted}, which usually means that a require somewhere failed; {\it Bad instruction}, which usually means that an assertion failed, and {\it Bad jump destination}, which can mean a number of things but the most likely meaning is that an array index was out of bounds. This is not helpful and detrimental to the transparency of the contracts. We believe that blockchain explorers should be able to report the reason for transaction failure when such reasons are programmed in the smart contract. The required data is available in the execution traces of the bytecodes. We would also encourage developers to make the reason for failure explicit in their code.

\subsection{Expose all callers of a smart contract}
On 1 Jan 2019, one of our sample contracts, {\normalsize \verb=EarthcryptToken= (See~\url{https://etherscan.io/address/0xe416b41a26661b00e2f41545eb7cef7995090aaf#code}) had received calls from 28,202 different addresses, of which 92\% via internal transactions and only 8\% via external transactions. The state of the contract therefore depends to a large extent on the changes made by the internal transactions that are not shown by blockchain explorers. About half the contracts of our random sample (659) receive calls via internal transactions. Current blockchain explorers only show these calls for the caller, and not for the callee. Given than there are about twice as many internal as external calls, and that many contracts make and receive internal calls, blockchain explorers should be able to trace callers {\it and} callees.

\subsection{Publish test boxes}
The Truffle framework makes it possible to package contracts, migration scripts, configurations, tests and more in a {\it Truffle box}. (See~\url{https://truffleframework.com/boxes}). The data is screened before a box is listed on the website, to ensure compatibility with the Truffle framework. Developers can choose from about 50 boxes as a starting point for a new development. We believe this to be a useful feature and we should like to propose that blockchain explorer publishes a {\it test box}, which contains the source of a smart contract, suitable migration and configuration scripts, and the test script that ContractVis has generated for the first $T$ historic transactions. The test script should be useful to the developer as a starting point for a regression test for the contract. The test script should also be useful to the user who is interested in experimenting with the contract, in the same way as we have experimented with Vitaluck.

\section{Discussion}
\label{sec:discussion}
We will now answer the research questions that we posed earlier as follows.

\textit{Q1:} In principle, a blockchain provides the perfect opportunity to study history by recreating it in considerable detail. This provides maximum transparency about the contracts that created the historic transactions. However, recreating history requires that the appropriate tools provide full details and that historic versions of the tools are available. Unfortunately, this is not always the case. Etherscan normally provides all the information necessary to replay the contracts. However, there are a number of minor issues, for example missing constructor arguments. We also found that older versions of the Solidity compiler do not work anymore, and that older versions of the {\normalsize \verb=ganache=} client are not available. We have reported these issues to the developers. In Section~\ref{sec:recommendations}, we provide five concrete suggestions for improvement of blockchain explorers such as Etherscan. Each suggestion improves a particular aspect of the transparency of smart contracts.

\textit{Q2:} We have been able to redeploy over 90\% of a sample of verified smart contracts on a minimal testing environment and we have been able to replay about 75\% of the first 50 historic transactions of these contracts. The termination status of the replayed transaction perfectly agrees with the termination status of the historic transaction. Where there is disagreement this is due to missing dependencies during the replay. We were expecting dependencies to have a bigger effect. It may be that our choice of replaying only the first $T=50$ transactions plays a role. For example, it could be the case that during the early stages of its life, a smart contract is setting up its administration and opens for business later. However, we have investigated to what extent our results are sensitive to the particular value of $T$ and found that above a certain value, the results are more or less stable.

\textit{Q3:} Our visualisations can automatically track the output of smart contract transactions in an intuitive manner. We have illustrated this in a case study of a lottery contract, confirming that using the time stamp of the current block as a source of randomness has issues. We have shown that visualisations with heat maps provide insight in the workings of a sample smart contract. The tooling automatically generates the heat maps. However, the user of the heat maps has to think of appropriate question to ask. For example: what would happen if the block times were a consecutive sequence? Would my lottery still be fair? Rerunning the transaction shows how the behaviour of the various runs differs, and helps to answer such what-if questions.

\section{Related work}
\label{sec:related}
Closest to our work is that of Cook et al~\cite{Cook2017}, who have developed DappGuard, a proof of concept tool that collects data from transactions of selected contracts, such as statistics on the gas usage per transaction and the value carried by transactions. Outliers, skewed distributions, and other irregularities are indicators of certain classes of attacks~\cite{Atzei2017}. This is similar to what we have shown for example in Figure~\ref{fig:Vitaluck_gasUsed}.

Anderson et al.~\cite{Anderson2016} analysed the bytecode of all 17,424 smart contracts available on the Ethereum block on 18 April 2016. During the early days of Ethereum 2,937 (i.e. 16.8\%) of the contracts were also similar to each other. We have used the same distance metric on our sample of smart contracts, but on the source rather than the binary. Comparing the source has the advantage that it is easier to explain the differences than on the binary. Heged\"us~\cite{Hegedus2019} has applied a number of well-known software metrics to the source code of 40,352 smart contracts. It seems that smart contracts score lower on key metrics that typical Object Oriented code. This might be an indication that the level of maturity of smart contract developers is lower than that of general OO developers.

ContractFuzzer~\cite{Jiang2018} is a fuzzing tool for smart contracts that uses abstract interpretation techniques to build the test oracles that generate the random inputs for the replay. Contractfuzzer has been tested on 9,960 verified smart contracts from Etherscan, of which about 1/3 could not be replayed because the authors had not considered replacing historic addresses.

Much related work deploys sophisticated techniques for model checking, static analysis, and abstract interpretation that we declared out of scope for our minimal testing environment. However, many of these techniques could be deployed to improve our minimal testing environment. Therefore we present an alphabetic selection of the most relevant tools.

Erasys~\cite{Zhou2018b} is a reverse engineering tool built to give users of smart contracts insight in contracts for which no source code is available. Compared to other bytecode, such as the Java Virtual Machine bytecode, the EVM bytecode is low level and therefore harder to reverse engineer. Erasys is able to uncover the functions in about 50\% of the contracts.

MadMax~\cite{Grech2018} is a tool that uses static analysis techniques to discover when smart contracts may run out of gas. Running out of gas is a clear indication that something is wrong with a transaction, hence an important topic of investigation. MadMax has been evaluated on the bytecode of over 1M smart contracts.

Maian~\cite{Nikolic2018} is a symbolic analysis tool designed to find trace vulnerabilities in smart contracts. The tool performs symbolic analysis of multiple execution traces of a contract to find safety and liveness issues. Maian has been tested on the bytecode of 970,898 smart contracts.

Osiris~\cite{Torres2018} is a tool that uses symbolic execution and taint analysis to discover integer overflow and underflows in smart contracts. This is another common cause for security problems. Osiris has been tested on 22,493 smart contracts from the Ethereum blockchain.

SASC~\cite{Zhou2018a} is a tool that looks for logical flaws in smart contracts. These are hard coded into a proof of concept tool. This approach does not require the programmer to specify patterns or policies, and is therefore easier to use but at the same time less powerful. SASC has been tested on 4,744 verified smart contracts from Etherscan.

Securify~\cite{Tsankov2018} is a code analysis tool that looks for specific code patterns known to be associated with security incidents. The patterns are coded by hand, using a powerful logic-based language. Securify has been evaluated using a sample of 24,594 smart contracts obtained from the Ethereum blockchain. The results indicate that typical issues, such as the DAO problem~\cite{Atzei2017} are indeed recognised by the tool.

SmartAnvil~\cite{Ducasse2019} is a toolkit designed to facilitate static analysis and bytecode reverse engineering of smart contract code. The main motivation for this tool is also to provide more insight in the functionality of smart contracts for which the source is unavailable.

teEther~\cite{Krupp2018} is a symbolic security checker that generates attacks on smart contracts with specific types of vulnerabilities. The tool has been tested on the bytecode of 38,757 unique smart contracts.

Zeus~\cite{Kalra2018} is a symbolic security checker. Where Securify requires developers to specify patterns that represent an issue, Zeus requires developers to specify policies that represent the avoidance of an issue. Zeus has been evaluated on the Securify dataset.

\section{Limitations}
\label{sec:limitations}
ContractVis is specific to Ethereum. There are hundreds of other flavours of smart contract systems. Some of these may be sufficiently similar to Ethereum such that the principles of our approach would also be applicable.

Contracts that have not been deployed yet do not have a public record of historic transactions that can be decompiled into a test. Developers of new contracts can use ContractVis with data from a test net.

The Truffle framework makes it possible to specify which version of the Solidity compiler to use, but it is not possible to specify the appropriate version of the Ethereum Virtual Machine. Since the specification of the EVM changes from time to time, particularly when hard forks are made, this is a limitation of our approach.

Our conclusions are valid for a random sample of about 10\% of a set of over 50,000 smart contracts. We have run the same experiments on several other, smaller random samples, all leading to similar results. Therefore we are confident that our results are representative for the collection of verified smart contracts on Etherscan.  Our results are probably not representative for the about 10M smart contracts also deployed on the Ethereum blockchain, for which the source code is not available on Etherscan. The developers of verified smart contracts made a special effort to upload their work onto Etherscan. It is therefore likely that these contracts were also developed with more care than other, unlisted, contracts. 

\section{Conclusions and future work}
\label{sec:conclusions}
We have provided a minimal testing environment for automated replay, visualisation, and analysis of the history of verified smart contracts.  We have been able to replay a random sample of 1120 verified smart contracts in isolation on this minimal replay environment. About 75\% of the transactions could be replayed accurately. The remaining 25\% could not be replayed because of missing dependencies. Many smart contracts call methods of other contracts. Since either the caller or the callee is absent in our minimal testing environment, the replayed transaction will fail. The relevance of this result is not in the fact that smart contracts have dependencies per se, but in the scale of the result: We had expected the effect of dependencies to be larger than 25\% failures. A manual inspection of a sample of tests revealed that our tests provide limited code coverage. This might explain the relatively low effect of dependencies on replay accuracy.

In a case study of a lottery contract, we have shown how a well-known issue, i.e. the lack of high-entropy source of randomness, can be visualised automatically and intuitively with heat maps. The heat maps also illustrate how the contract leaks crucial information via the blockchain. The ability to replay the contract multiple times, while varying the environmental parameters is key to achieving intuitive visualisations. A source code audit would have come to the same conclusion about poor design of the lottery contract, but we believe that our heat maps are more intuitive.

In general, verified smart contracts are less transparent than we would have expected. About half the contracts in the sample use special variables, and have therefore a potentially large range of different behaviours. About half the contracts in the sample are clients that use other contracts as servers, and a minority are server contracts. To assess the transparency of the client, users have to be able to inspect the server and vice versa. Current blockchain explorers provide insufficient support for such inspections. About 10\% of the sample contains ad-hoc encodings of addresses, which current blockchain explorers do not decode, thus reducing transparency.

Most transactions that fail programmatically on a {\normalsize \verb=require(.)=} or {\normalsize \verb=assert(.)=}, do so without stating a reason. Developers could improve this significantly with little effort, by indicating the reason for failure, for example {\normalsize \verb=require(.,"method cannot be called at this time")=} rather than {\normalsize \verb=require(.)=}.

Etherscan and other blockchain explorers provide a useful and free service to the community. However, there is room for improvement. We have made five recommendations in Section~\ref{sec:recommendations} that we hope will be implemented.

Future work includes studying ways to lift some of the limitations of ContractVis. Firstly, it would be interesting to investigate how large the dependencies of a contract are and how these could be incorporated in a testing environment. Secondly, it may be possible to use a decompiler such as Erasys~\cite{Zhou2018b} to obtain source code that is acceptable to {\normalsize \verb=truffle test=}. Then ContractVis could also be used for contracts without source code.

\section*{Acknowledgments}
We thank Maarten Everts, Dani\"el Reijsbergen and Richard Schumi for their comments on a draft of this paper. We have asked 16 technical questions on Etherscan and we have opened 14 issues on the Truffle and Ethereum repositories. We thank the developers for their replies, which usually arrived swiftly.

The generated {\normalsize \verb=truffle test=} scripts and some of the details of the analysis results for our sample of 1,120 verified smart contracts are available on GitHub (See \url{https://github.com/pieterhartel/Truffle-tests-for-free}).

\bibliography{darkweb_refs}

\end{document}